# Giant enhancement of cryogenic thermopower by polar structural instability in the pressurized semimetal MoTe$_2$

Hidefumi Takahashi[1,2], Kento Hasegawa[1], Tomoki Akiba[1], Hideaki Sakai[3,4], Mohammad Saeed Bahramy[1,5], and Shintaro Ishiwata[1,2,4]

[1]Quantum-Phase Electronics Center (QPEC) and Department of Applied Physics, The University of Tokyo, Tokyo 113-8656, Japan

[2]Division of Materials Physics, Graduate School of Engineering Science, Osaka University, Toyonaka, Osaka 560-8531, Japan

[3]Department of Physics, Osaka University, Toyonaka, Osaka 560-0043, Japan

[4]PRESTO, Japan Science and Technology Agency, Chiyoda, Tokyo 102-8666, Japan

[5]RIKEN Center for Emergent Matter Science (CEMS), Wako, Saitama 351-0198, Japan

**We found that a high-mobility semimetal 1T′-MoTe$_2$ shows a significant pressure-dependent change in the cryogenic thermopower in the vicinity of the critical pressure, where the polar structural transition disappears. With the application of a high pressure of 0.75 GPa, while the resistivity becomes as low as 10 μΩcm, thermopower reached the maximum value of 60 μVK$^{-1}$ at 25 K, leading to a giant thermoelectric power factor of 300 μWK$^{-2}$cm$^{-1}$. Based on semi-quantitative analyses, the origin of this behavior is discussed in terms of inelastic electron–phonon scattering enhanced by the softening of zone-center phonon modes associated with the polar structural instability.**

Enhancement of thermoelectric effects at low temperatures is demanded for the development of cryogenic Peltier coolers. From the viewpoint of the power output of these devices, it is necessary to find the material with high power factor (= $S^2\rho^{-1}$), where $S$ and $\rho$ denote thermopower and resistivity, respectively. In the cryogenic temperatures below 50 K, strongly correlated electron systems such as cobalt oxides and heavy-fermion compounds [1–5] were found to show much larger power factors than those for high-mobility semiconductors and semimetals typified by Bi$_2$Te$_3$ (see Fig. 4) [6-10]. Since thermopower is the measure of the entropy normalized per charge carrier, the unusually high power factors originated from the high thermopower in these compounds were discussed in terms of the large entropy of spin and orbital degrees of freedom.

As well as the spin and orbital entropy, the phonon entropy has a potential for the enhancement of thermopower through electron–phonon scattering [11,12]. A typical example is the phonon-drag effect observed in the clean system with long mean free path phonons [12]. However, since the enhancement of thermopower through the phonon-drag effect requires a high thermal conductivity $\kappa$, the improvement of the dimensionless thermoelectric figure of merit $zT$ ($=S^2 T \rho^{-1} \kappa^{-1}$) is less promising. Here we report the discovery of gigantic cryogenic power factor in semimetallic 1T′-MoTe$_2$ at low temperatures and high pressures where thermopower is enhanced through the inelastic electron-phonon scattering near the polar-nonpolar structural phase boundary.

Recently, 1T′-MoTe$_2$, which has been intensively studied as a Weyl semimetal candidate [13,14], was found to show an anomalous enhancement of thermopower at low temperatures as Mo is partly substituted by Nb [15]. This anomalous enhancement is seemingly associated with the polar structural fluctuation near the critical concentration, where the polar structural transition temperature $T_s$ is decreased to be zero. Note that the phonon-drag effect can be ruled out as an origin of the anomalous enhancement of thermopower, because the thermal conductivity of 1T′-Mo$_{1-x}$Nb$_x$Te$_2$ remains small at low temperatures (see Ref. [15]). However, since the chemical substitution accompanies a change in the chemical potential and an increase in the impurity scattering, the relationship between the enhanced thermopower and the polar fluctuation in Nb-doped 1T′-MoTe$_2$ remains unclear. In addition, the chemical disorder hampers the improvement of power factor by increasing the electrical resistivity. In contrast, the application of external pressure also suppresses $T_s$ without inducing major shifts in the chemical potential and chemical disorder [16,17]. In this work, we successfully observed a gigantic power factor in single-crystalline samples of 1T′-MoTe$_2$ at low temperatures and high pressures (the measurement setup is depicted in Supplemental Material Fig. S1 [18]), which provides an important clue for understanding the role of the polar structural instability on the enhancement of the cryogenic thermopower. Based on semi-quantitative analysis, we discuss the possible role of inelastic electron–phonon scattering involving the zone-center soft phonons in the remarkable enhancement of thermopower, providing a promising strategy for exploring new thermoelectric materials.

Single crystals of 1T′-MoTe$_2$ were prepared by the flux method with NaCl. The quality of the crystals in terms of the residual resistivity ratio (RRR) was found to be significantly better in the case of the flux method (RRR ~ 300) as compared to the case of the chemical vapor transport method (RRR ~ 60) reported in the previous study [17]. The typical dimensions of a single crystal were $4 \times 1 \times 0.1$ mm$^3$. Electrical resistivity ($\rho$) and thermopower ($S$) were measured from 300 to 2 K using a Physical Property Measurement System (Quantum Design, Inc.). Pressure was applied using a piston-cylinder clamp cell composed of BeCu. Daphne oil 7373 was used as the pressure-transmitting medium.

To clarify the origin of the remarkable enhancement of thermopower, we firstly studied the pressure-dependence of electrical resistivity. The inset of Fig. 1b shows the in-plane resistivity ($\rho$) as a function of temperature ($T$) for 1T′-MoTe$_2$ at selected pressures ($P$). As indicated by the arrows, the anomaly corresponding to the structural transition temperature ($T_s$) shifted to lower temperature with increasing pressure, becoming smeared out above 0.75 GPa (The anomaly at $T_s$ is clearly discernible in the temperature derivative of ρ as shown in Supplemental Material Fig. S2 [18]). The structural phase diagram as a function of $T$ and $P$ is summarized in Fig. 1b. Extrapolation of $T_s$ as a function of $P$ revealed that the polar structural transition disappeared at the critical pressure of $P_c ≈ 0.9$ GPa. It should be noted here that the pressure dependence of ρ changes smoothly at low temperatures around $P_c$, in contrast to that of thermopower, which will be discussed in detail below.

To gain further insight into the influence of pressure on the band structure, we measured the Hall resistivity (ρ$_{yx}$) to evaluate the Hall coefficient ($R_H$) at selected pressures and a temperature of 5 K, as shown in Fig. 2a (the results for ρ$_{yx}$ above 5 K are presented in Supplemental Material Fig. S3 [18]). The Hall resistivity was negatively and linearly dependent on the magnetic field, which holds for the single carrier model ($n = 1/eR_H$), where $n$ represents the carrier concentration of electrons. The slopes of the lines decrease with increasing pressure in accordance with the previous study, which suggests that the carrier concentration increased upon the application of pressure [19]. Figure 2b and the inset of Fig. 2c show the temperature dependence of ρ and $R_H$, respectively, below 30 K at various pressures, revealing a qualitatively similar pressure effect; ρ($T$) and $R_H$($T$) varied slightly in the polar phase and were substantially suppressed in the nonpolar phase upon the application of pressure. Figure 2c shows the variation of $R_H$ and ρ as a function of pressure at 5 K (The data below 30 K are plotted in Supplemental Material Fig. S4 [18]). Since the carrier concentration evaluated by the single-carrier model at ambient pressure (~ $4 \times 10^{20}$ cm$^{-3}$) was almost the same as that precisely evaluated by the two-carrier model (~ $2 \times 10^{20}$ cm$^{-3}$; see Figs. S5 and S6 for more details), the effect of pressure on $R_H$ apparently reflects the variation in the carrier concentration [20]. The validity of the single carrier model was also supported by the experimental fact that the mobility μ of electrons (~ $2.5 \times 10^4$ cm$^2$V$^{-1}$s$^{-1}$) is much higher than that of holes (~ $0.5 \times 10^4$ cm$^2$V$^{-1}$s$^{-1}$; see Fig. S6). It should be noted that the pressure dependence of the resistivity can be scaled with that of $R_H$, indicating that the pressure dependence of the resistivity is dominated by the change in carrier concentration rather than the scattering rate.

Figure 3a presents the temperature dependence of thermopower $S$ at various pressures. In contrast to the resistivity, a pronounced pressure effect was observed in the vicinity of $P_c$ at low temperatures. At ambient pressure ($P ≈ 0$ GPa), the value of $S$ underwent sign changes at 125 and 250 K, reflecting the semimetallic band structure. Furthermore, a significant positive peak was discernible below 50 K. Upon increasing the pressure to 0.75 GPa (i.e., to just below $P_c$), a positive peak in the $S(T)$ curves developed at 35 K and reached the relatively high value of 60 μVK$^{-1}$. Upon further increasing the pressure to 1 GPa and above, the magnitude of the peak decreased. The subtle influence of the pressure on *S* at room temperature indicates the negligible change of the chemical potential upon the application of pressure. A contour plot of *S* is superimposed on the *P*–*T* phase diagram in Fig. 3b. The evolution of *S*(*T*) at low temperatures was clearly observed just below $P_c$, strongly suggesting that destabilization of the polar structure was responsible for the enhancement of the peak in the *S*(*T*) curves. This anomalous feature was absent for the resistivity as a function of temperature. Thus, the carrier concentration and relaxation time are presumably less dependent on temperature and pressure even in the vicinity of $P_c$. As a possible origin of the low-temperature peak in the *S*(*T*) curves, the phonon-drag effect is typically considered. In general, the phonon drag effect is often found in high purity crystals, which reflects the fact that the enhancement of the thermopower by this effect is proportional to the phonon mean free path [21]. However, this effect should be absent in this system because polycrystalline samples of Nb-doped 1T′-MoTe$_2$, in which the phonon mean free path is expected to be much shorter than that in the single-crystalline sample of 1T′-MoTe$_2$, also exhibited a similar peak in the *S*(*T*) curves as discussed in Ref. [15].

Here, we discuss the possible origin of the anomalous pressure-induced enhancement of thermopower by considering the change in the band structure and the scattering mechanism. The diffusion thermopower based on the Mott formula, $S_M$, is determined by the logarithmic energy derivative of the electrical conductivity tensor. For a three-dimensional free-electron system, we obtain

$$S_M = -\frac{\pi^2 k_B^2 T}{e}\left[m^*(3\pi^2 n \hbar^3)^{-\frac{2}{3}} + \frac{1}{3}\frac{\partial \ln \tau(\varepsilon)}{\partial \varepsilon}\right]_{\varepsilon=\varepsilon_F}, \qquad (1)$$

where $m^*$, $n$, and $\tau(\varepsilon)$ denote the electron effective mass, carrier concentration, and relaxation time of the conduction electrons, respectively. This expression indicates that $S_M$ is composed of the band contribution $S_{M1}$ (first term) and the scattering contribution $S_{M2}$ (second term). The band contribution of Eq. (1) essentially depends on *n* and $m^*$, the latter of which can be estimated from the temperature dependence of the Shubnikov–de Haas oscillation (see Supplemental Material Figs. S7 and S8 [18]). As pressure was increased from ambient pressure to 0.5 GPa, $m^*$ for electron carriers increased from 0.57$m_0$ to 0.72$m_0$, where $m_0$ is the bare electron mass, while the carrier concentration remained almost constant. If Eq. (1) holds for the thermopower of the present system, the increase ratio of the pressure-induced enhancement of the thermopower through the change in $m^*$ is expected to be 125%. This ratio is, however, substantially lower than the experimentally observed increase ratio (170%). In addition, *S*/*T* at the lowest temperature, which is proportional to $m^*$ for metallic systems, monotonically decreased, and even underwent a change in sign with increasing pressure to afford a negative minimum (see Supplemental Material Figs. S9 and S10). Therefore, it is reasonable to conclude that the anomalous enhancement of thermopower at low temperatures cannot be described by Eq. (1) assuming the relaxation time approximation derived from the Mott formula, while the pressure-induced enhancement of $m^*$ signifies a subtle band modification.

To quantitatively analyze the band contribution to the thermopower, we calculated $S_M$ for the orthorhombic phase at ambient pressure by adopting the experimental values of the carrier concentration ($n = 1/eR_H$) and $m^*$ for Eq. (1) and neglecting the scattering contribution $S_{M2}$. As shown in Fig. 3c, $S_M$ (= $S_{M1}$) did not reproduce the low-temperature peak observed in the experimental result. The temperature dependence of thermopower has also been estimated from first-principles band-structure calculations for the orthorhombic phase [15]. The calculated thermopower $S_{FP}$ takes full account of the band contribution, while the scattering contribution is excluded by adopting the constant-$\tau$ approximation. As shown in Fig. 3c, $S_M$ is fairly similar to $S_{FP}$, supporting our assumption that $S_M$ corresponds to the band contribution for the total thermopower.

Next, we consider the scattering contribution as a possible origin of the anomalous enhancement of thermopower. A large scattering contribution due to a highly dispersive $\tau(\varepsilon)$ was recently reported in strongly correlated systems and materials with continuous magnetic and structural transitions associated with electron scattering by local spins (Kondo scattering) and critical fluctuation due to the phase transition, respectively [22–25]. These scattering mechanisms, however, usually affect $\rho(T)$, which is not the case for 1T′-MoTe$_2$ as it exhibited no anomaly in $\rho(T)$ in the vicinity of $P_c$. The absence of an anomaly in $\rho(T)$ also indicates that domain boundary scattering, which is supposed to be enhanced near $T_s$, makes a minor contribution to the enhancement of thermopower.

As an alternative scattering mechanism for the anomalous features of thermopower, we consider inelastic electron–phonon scattering. Although inelastic scattering typically necessitates only a slight correction to the Mott formula for thermopower [26], the phonons related to structural instability may significantly influence thermopower through band renormalization or a characteristic scattering process. Here, we consider an inelastic-scattering model including the "vertical" scattering process reported in Ref. [27] to describe the anomalous enhancement of thermopower as follows;

$$S = S_0(t) + S_1(t)\frac{k_B T}{\varepsilon_F} , \qquad (2)$$

where

$$S_0(t) = \frac{\pi^2}{2}\left(\frac{k_B}{e}\right)\frac{P_{12}(t)}{P_{22}(t)} , \qquad (3)$$

$$S_1(t) = -\frac{\pi^2}{2}\left(\frac{k_B}{e}\right)\left\{1 + \frac{\pi^2}{3}\frac{P_{11}(t)}{P_{22}(t)}\right\} . \qquad (4)$$

The term $S_0(t)$ is effective only if an inelastic contribution to the scattering is taken into account. To

evaluate the value of $S$, we need to calculate the scattering matrix elements expressed as follows:

$$P_{11}(t) = F_5(t), \quad (5)$$

$$P_{12}(t)P_{21}(t) = \frac{1}{2}F_6(t), \quad (6)$$

$$P_{22}(t) = F_7(t) - P_-(t), \quad (7)$$

with

$$P_-(t) = \frac{\varepsilon_s}{k_B T}\left[\frac{1}{2}F_6(t) - \frac{\varepsilon_F}{k_B T}P_{11}(t)\right] \quad (8)$$

and $t = T/\Theta_D$ ($\Theta_D$ is Debye temperature). The generalized ($n$ = 5, 6, 7) Bloch (-Grüneisen) functions

$$F_n = \int_0^{1/t}\left[\frac{y^n dy}{(e^y-1)(1-e^{-y})}\right] \quad (9)$$

with the material constants $\Theta_D$, $\varepsilon_F$, and $\varepsilon_s$ are sufficient for the description of the scattering. Here, $\varepsilon_F$ is the Fermi energy, and $\varepsilon_s$ is $2m^*v_s^2$, which is a material parameter with energy scale interrelating the electron and phonon systems: the sound velocity $v_s$ determines the "stiffness" of the materials (see the Supplemental Material for more details [18]). Note that the "vertical" scattering process accompanies the change in the kinetic energy of conduction electrons but is unrelated to their momentum. Since the thermal smearing of the Fermi distribution function allows the "vertical" scattering process, this process can contribute to the thermopower caused by thermal electron diffusion. This model indicates that thermopower as a function of temperature $S(T)$ is distinct from that of the Mott formula well below the Debye temperature of $\Theta_D \approx 180$ K [17]; the inelastic scattering term $S_0(t)$ [$\approx S(T) - S_M(T)$] is additive to the Mott formula. As shown in Fig. 3c, the structure of the low-temperature peak can be reproduced with reasonable parameter values based on the experimental results at ambient pressure [$\Theta_D \approx 180$ K [17], and $\varepsilon_F = 350$ meV, which is estimated from the carrier concentration $n$ (= $4\times10^{20}$ cm$^{-3}$) and $m^*$ (= 0.58 $m_0$) with assuming the three dimensional parabolic band ($\varepsilon_F = \hbar^2(3\pi^2 n)^{2/3}/2m^*$)]. We note here that the energetic parameter of $\varepsilon_s = m^*v_p^2$, where $v_p$ is the phonon velocity, determines the low-temperature value of $S(T)$ additively enhanced by inelastic electron–phonon scattering. Thus, the anomalous enhancement of thermopower is attributable to the suppression of $\varepsilon_s$ through the decrease in $v_p$, which possibly originates from the polar structural instability. As shown above, the inelastic electron–phonon scattering model reproduces characteristic features such as the low-temperature positive peak in the $S(T)$ curves. It should be noted here that $S(T)$ shows a nonmonotonic temperature dependence below 30 K above 1.0 GPa (see Fig. 3a and Supplemental Material Fig. S10 [18]). This behavior presumably reflects the fact that, whereas the electron carriers contributes negatively to $S$ especially at low temperatures below 30 K, the inelastic electron-phonon scattering yields positive peak in $S$ around 30 K [28]. Since the inelastic "vertical" scattering is unrelated to their momentum, this model is consistent with the fact that no significant anomaly was observed for $\rho(T)$ under high pressures in the vicinity of $P_c$.

For inelastic "vertical" scattering, low-energy phonon modes near the zone center presumably play an important role. In fact, Raman spectroscopy and density functional theory calculations of phonon dispersion for 1T′-MoTe$_2$ revealed the existence of an extremely low energy optical phonon mode (~9.2 cm$^{-1}$ for the orthorhombic polar phase and ~15.3 cm$^{-1}$ for the monoclinic nonpolar phase) at the zone center [29]. The inelastic scattering between the electrons and the low energy phonons enhances the thermopower at low temperature in both the orthorhombic polar phase at ambient pressure and the monoclinic nonpolar phase at high pressure (> 1 GPa). In the vicinity of $P_c$, softening of the zone-center phonon mode associated with the polar structural transition presumably reduces the velocity of phonons involved in inelastic electron–phonon scattering, thus representing a source of the anomalous enhancement of thermopower, that is, the external pressure decreases the structural transition temperature, and around $P_c$ the energetic parameter of $\varepsilon_s$ is reduced at low temperature through the phonon softening, resulting in the enhancement of the thermopower. In contrast, 1T′-WTe$_2$, which possesses the same crystal structure and similar band structures as 1T′-MoTe$_2$, exhibits no anomaly in its $S(T)$ curves at low temperature[15], reflecting the much higher energy of its zone-center optical mode (~45 cm$^{-1}$) [30].

Owing to its greatly enhanced thermopower of 60 μVK$^{-1}$ in combination with its rather low resistivity of 10 μΩcm at 25 K and 0.75 GPa, 1T′-MoTe$_2$ exhibits a maximum power factor of ~300 μWK$^{-2}$cm$^{-1}$ at low temperatures, as shown in Fig. 4. This value even surpasses that of the heavy-fermion system YbAgCu$_4$, which is considered one of the best thermoelectric materials for cryogenic applications due to the electron diffusion process [1]. It should be noted here that an exceptionally large thermopower of 1-45 mVK$^{-1}$ and a high power factor of 50~8000 μWK$^{-2}$cm$^{-1}$ were observed for FeSb$_2$ at low temperatures [31,32]. However, the maximum value of the expected dimensionless figure of merit $zT$ for FeSb$_2$ (~ 0.005 at 15 K) is much smaller than that for 1T′-MoTe$_2$ (~ 0.038 at 25 K), where the thermal conductivity is assumed to be constant (~ 20 W/mK at 25 K) against pressure below $P_c$. In addition, reflecting the phonon-drag mechanism, the thermopower of FeSb$_2$ can be largely influenced by extrinsic factors such as the sample size and the impurity or domain boundary scattering [20,33]. For instance, the thermopower of the polycrystalline sample of FeSb$_2$ is 300 μV/K, which is significantly smaller than that of the single crystalline sample. On the other hand, the thermopower of 1T′-MoTe$_2$ is free from the extrinsic factors, indicating the predominance of the electron diffusion mechanism with the inelastic "vertical" electron–phonon scattering. Furthermore, 1T′-MoTe$_2$ with rather small effective mass ($m^*$) is an ideal thermoelectric material in terms of this scattering mechanism, since the enhancement of the low-temperature peak in $S(T)$ discussed in Fig. 3c is expected to be larger as $m^*$ becomes smaller. This work demonstrates that a high-mobility metallic system with soft phonon modes at zone center is a promising candidate for high efficiency thermoelectric applications at cryogenic temperatures.

dependence of the Shubnikov–de Haas oscillation, thermopower at low temperatures, description of the inelastic-scatterimg model.

**Acknowledgements**

The authors thank I, Terasaki, Y. Tokura, J. Fujioka, and K. Ishizaka for useful comments. This study was supported in part by KAKENHI (Grant No. 17H01195), JST PRESTO Hyper-nano-space Design toward Innovative Functionality (JPMJPR1412), the Asahi Glass Foundation, and the Mazda Foundation.

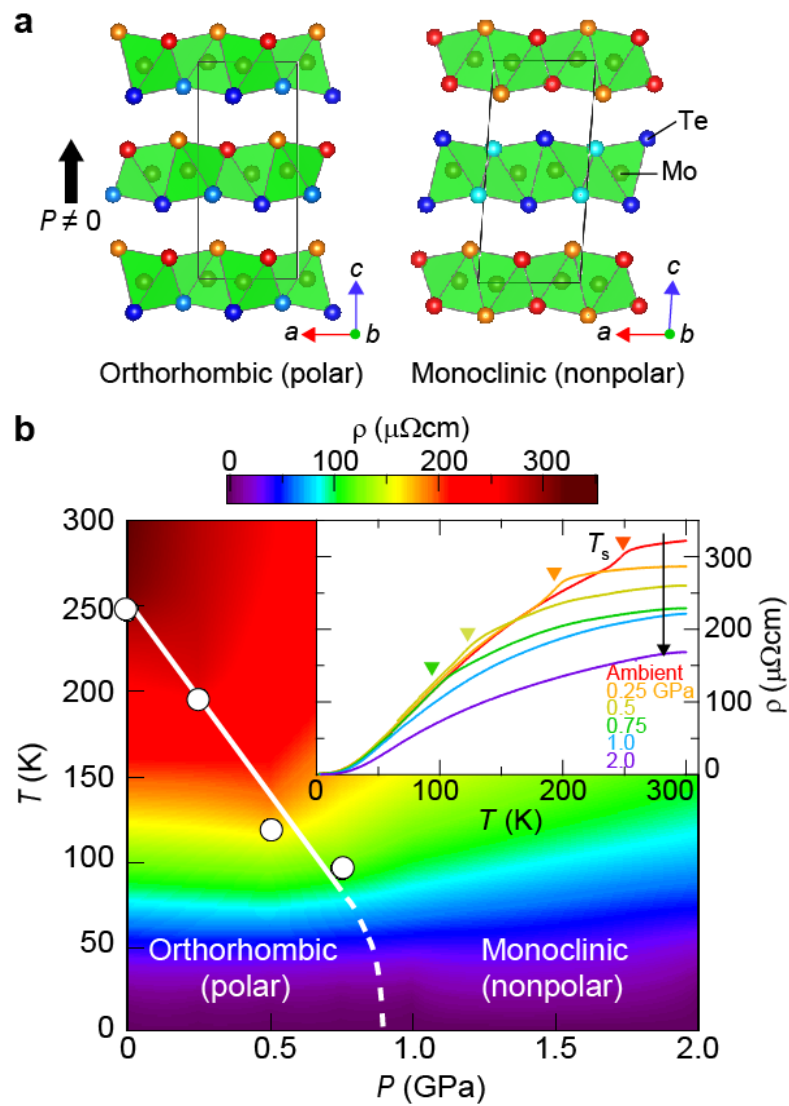

Figure 1. **Crystal structures of MoTe$_2$ and influence of pressure on resistivity ρ. a**, Schematic representations of the orthorhombic polar structure (T$_d$ phase) and monoclinic nonpolar structure (1T′ phase) of MoTe$_2$. The polar direction is parallel to the *c* axis as indicated by the black arrow. **b,** Contour plot of ρ as a function of temperature *T* and pressure *P*. The inset shows the temperature dependence of ρ at pressures of up to 2 GPa.

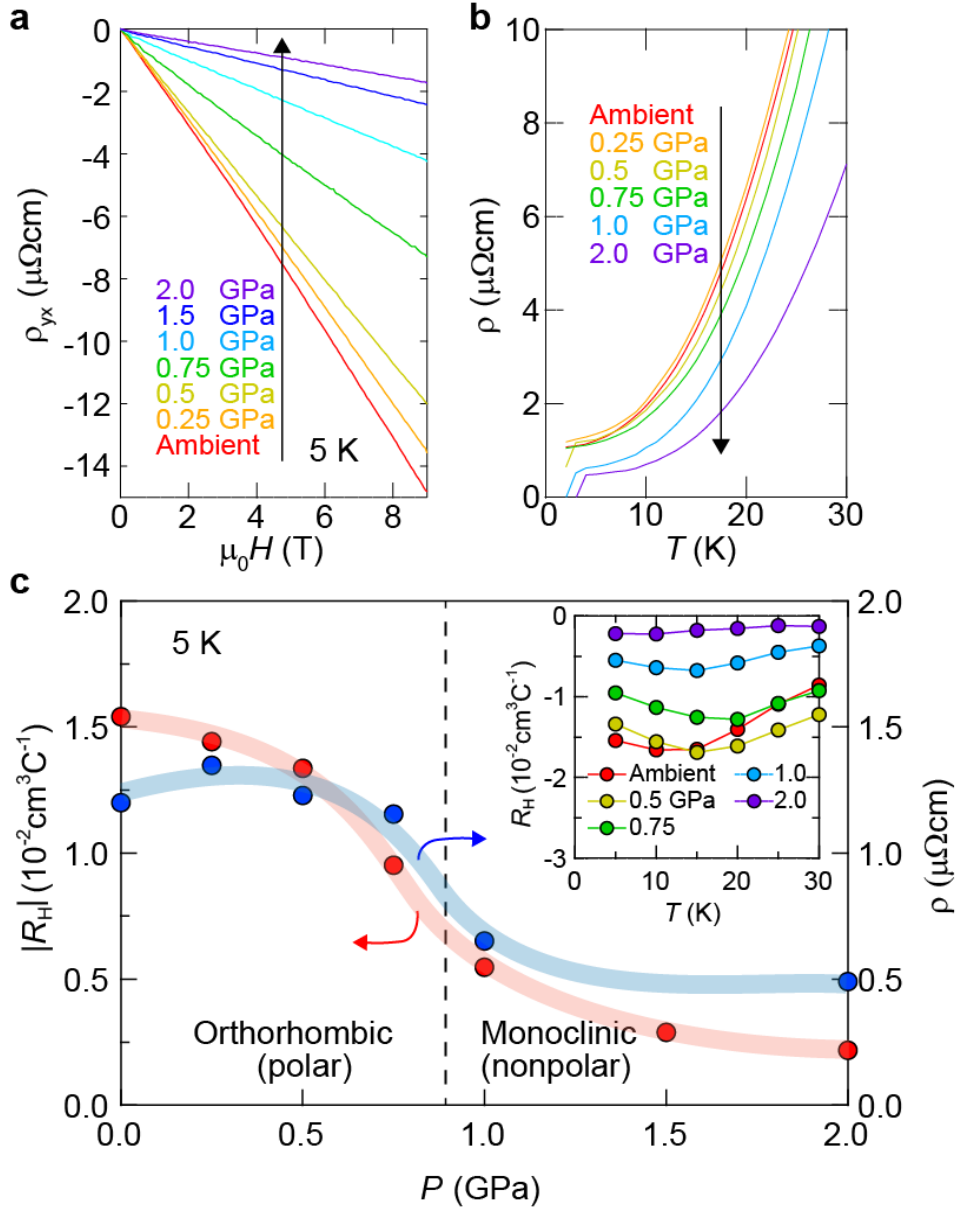

Figure 2. **Comparison between the pressure dependence of ρ and the Hall coefficient $R_H$. a,** Magnetic field dependence of Hall resistivity $\rho_{yx}$ at various pressures and a temperature of 5 K. **b,** Temperature dependence of ρ at various pressures below 30 K. **c,** Pressure dependence of ρ and $R_H$ at 5 K. The inset shows the temperature dependence of $R_H$ at various pressures below 30 K.

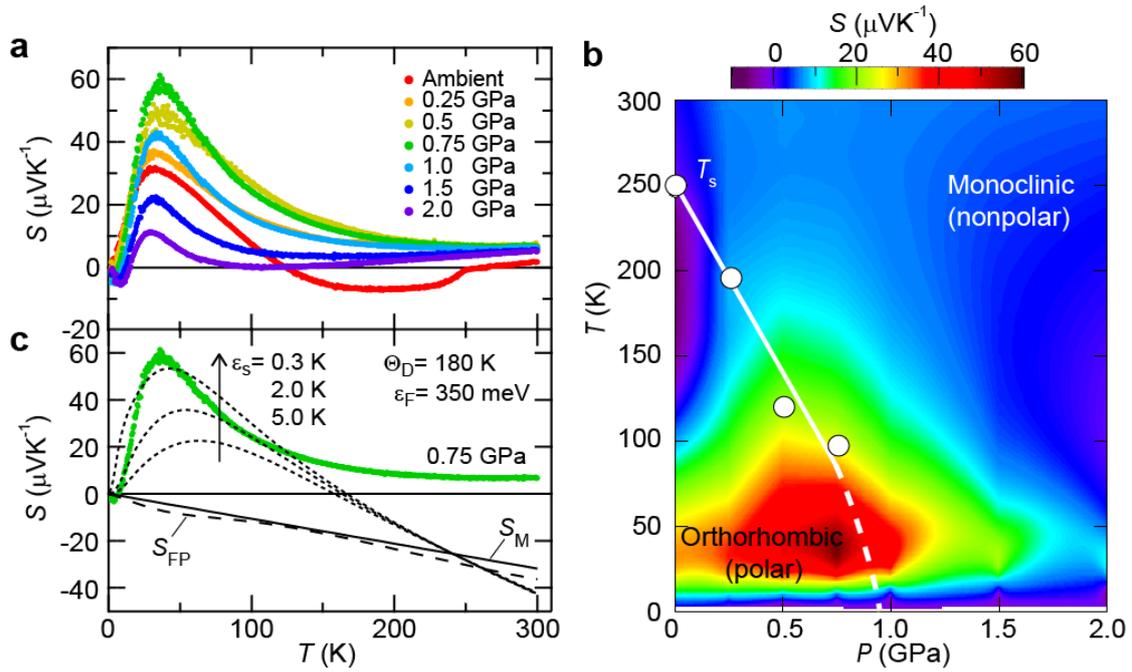

Figure 3. **Influence of pressure on thermopower $S$ and theoretical models of the enhanced $S$. a,** Temperature dependence of $S$ at various pressures of up to 2 GPa. **b,** Contour plot of $S$ as a function of temperature $T$ and pressure $P$. The structural phase diagram is overlaid on the plot. The circles denote the transition temperatures $T_s$ determined by $\rho$. A greatly enhanced $S$ is observed at low temperature in the vicinity of the critical pressure $P_c$, where the polar structural transition disappears. **c,** Comparison between the experimentally observed $S(T)$ curve (green) and the three theoretically calculated $S(T)$ curves (black); $S_{FP}$ (dashed line) was obtained by first-principles band calculations, $S_M$ (solid line) was calculated using Eq. (1) with experimentally obtained values of $m^*$ and $n$, and the theoretical $S(T)$ curves plotted as dotted lines were calculated using the inelastic electron–phonon scattering model. The Debye temperature of $\Theta_D = 180$ K was taken from Ref. [17] and the Fermi energy $\varepsilon_F$ is an arbitrary fitting parameter (see the Supplementary Information for details).

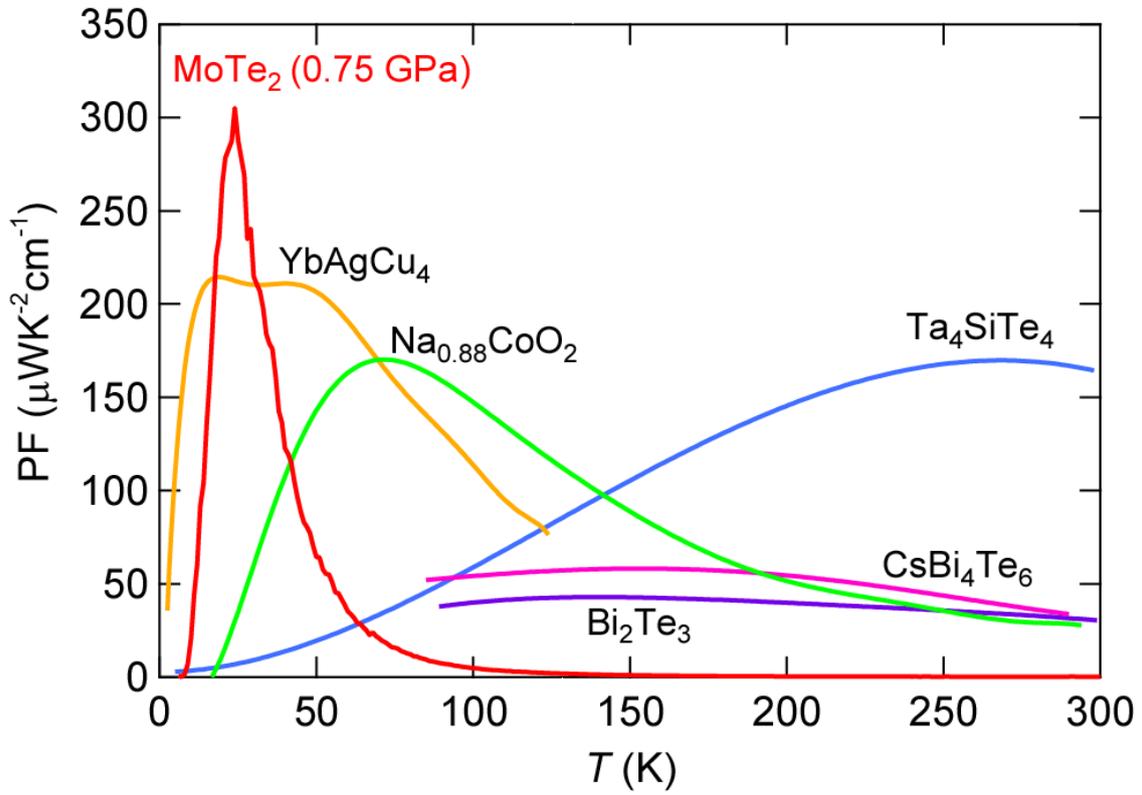

Figure 4. **Power factors as a function of temperature**. The power factors (PFs) of various thermoelectric materials below 300 K are plotted. A gigantic PF of ~300 μWK$^{-2}$cm$^{-1}$ can be observed for MoTe$_2$ in the vicinity of the critical pressure (~0.75 GPa). All data except for MoTe$_2$ are taken from Refs. [6-9,34,35].